\documentstyle[amsmath,amssymb,graphicx]{article}

\def\be{\begin{eqnarray}}
\def\ee{\end{eqnarray}}

\def\bfig{\begin{figure}[H] }
\def\efig{\end{figure}}

\def\bc{\begin{center}}
\def\ec{\end{center}}

\def\tr{{\rm tr}\,}
\def\Tr{{\rm Tr}\,}

\hoffset= - 1.0in         

\title{{\bf Finalizing the proof of AGT relations with the help of the generalized Jack polynomials} \vspace{.2cm}}

\author{ {\bf A.Morozov}\thanks{{\small
{\it ITEP, Moscow, Russia} }; morozov@itep.ru} , {\bf
A.Smirnov }\thanks{{\small {\it Department of Mathematics, Columbia University, New York USA} }; asmirnov@math.columbia.edu}\date{ }}

\begin{document}
\date{}
 \maketitle

\vspace{-6.0cm}

\begin{center}
\hfill ITEP/TH-25/13\\
\end{center}

\vspace{3.5cm}
\bigskip

\bigskip
\bigskip

\centerline{ABSTRACT}

\bigskip

{\footnotesize
Original proofs of the AGT relations with the help of the Hubbard-Stratanovich
duality of the modified Dotsenko-Fateev matrix model
did not work for $\beta\neq 1$, because Nekrasov functions were not
properly reproduced by Selberg-Kadell integrals of Jack polynomials.
We demonstrate that if the generalized Jack polynomials, depending on
the $N$-ples of Young diagrams from the very beginning, are used instead
of the $N$-linear combinations of ordinary Jacks, this resolves the problem.
Such polynomials naturally arise as special elements in the equivariant
cohomologies of the $GL(N)$-instanton moduli spaces, and this also establishes
connection to alternative ABBFLT approach to the AGT relations,
studying the action of chiral algebras on the instanton moduli spaces.
In this paper we describe a complete proof of AGT in the simple case
of $GL(2)$ ($N=2$) Yang-Mills theory, i.e. the 4-point spherical conformal block
of the Virasoro algebra. }


\bigskip

\begin{small}
\section{Introduction}
AGT relations \cite{AGT1}-\cite{AGT2} identify two {\it a priori} different classes of theories:
$2d$ conformal theories \cite{CFT1}-\cite{CFT3}
and instanton calculus in multi-dimensional Yang-Mills \cite{LMNS}.
It is a deep and far-going generalization of the Seiberg-Witten theory \cite{SW1}-\cite{SW2},
based on the insight into the quasiclassical physics of branes
in $M$-theory \cite{hz}-\cite{Ga} and related integrability properties \cite{SW3}-\cite{Nef} and \cite{ZZ}-\cite{ZOLS4}.
Closer to the Earth, at the moment there are two technical approaches to
study and prove these kinds of relations, based on embedding the two subjects
into something more general -- but not as big as the entire string or $M$-theory.

The first approach \cite{MMS1}-\cite{MMS} utilizes the free field representation of conformal block: this gives the integral
representation of conformal block in the form of
Dotsenko-Fateev (DF) integrals \cite{DF1}-\cite{DF4}. This approach leads to expansion of $SU(N)$ conformal blocks in series whose coefficients
are represented by $A_{N-1}$-Selberg integrals of Jack polynomials \cite{ZM}. However, as was stressed in \cite{MMS,MMSS}, Selberg averages of Jack polynomials reproduce the coefficients of the Nekrasov partition function only for central charge $c=1$. In general case, these averages are not factorizing to linear terms (as Nekrasov coefficients) and the expansion of conformal block does not have a form of Nekrasov partition function.

The second approach  \cite{Alba1}-\cite{Y3} exploits Nakajima's results
on the geometry of the instanton moduli spaces. The explicit form of the coefficients in the expansion of the conformal block
depends on the choice of basis of the intermediate states. As was noted in the series of papers  \cite{Alba1}-\cite{AlbaLast}, the most natural basis is given by the classes of the fixed points in the equivariant cohomology of the instanton moduli spaces (due to  Nakajima \cite{Nak1}-\cite{Nak6}, the equivariant cohomology of instanton moduli space is identified with the Fock space in CFT and the fixed point classes represent a natural basis of this space). In this basis, the coefficients of the conformal block coincide precisely with the coefficients of the Nekrasov function, which gives the proof of AGT relation.

In this paper we unify these seemingly different approaches and resolve $c\neq 1$
problem of the first of them:
the missing detail is a far-going generalization of the Kadell's formulae for
Selberg averages of two Jack polynomials.
We introduce two bases (dual to each other) in the Fock space, which coincide with the bases used in \cite{Alba1}
after bosonization of the Virasoro operators. We call these special polynomials \textit{generalized Jack polynomials}.
The expansion of the conformal block in the Dotsenko-Fateev representation
leads to Selberg integrals of the generalized Jack polynomials which are
\textbf{completely factorized to linear multiples and coincide with  coefficients of Nekrasov function for arbitrary choice of the central charge $c$!}

Moreover, in the limit $c=1$ the generalized Jack polynomials are reduced to a product of Schur polynomials $J_{\mu \nu} =s_\mu s_\nu$ and we obtain the results of \cite{MMS}.
Thus, we extend the proof \cite{MMS}
of AGT relations from $c=1$ to arbitrary $\beta$-deformation \cite{betadefo}.
Generalizations
to higher-rank gauge groups, to $5$ dimensions {\it a la} \cite{MMSS}
and in other directions  seem straightforward, since only the universal matrix-model technique is really needed.

This paper is organized as follows: in section \ref{agthb}
we briefly describe the approach to AGT relation based on
Dotsenko-Fateev representation of the conformal blocks.
In section \ref{basch} we discuss the choice of the basis for the intermediate states in the conformal block and remind the $\beta\neq 1$ problem of \cite{MMS}.
In the next section \ref{levone} we give a simple pedagogical exposition of our results in the simplest case of the generalized Jack polynomials at level one.
In section \ref{hamilt} we give a definition of the generalized Jack polynomials as eigenfunctions of hamiltonians defining some integrable systems.
The main formulae for the Selberg averages of the generalized Jack polynomials are given in section \ref{saver}. Using these formulae in section \ref{sproof} we give a proof of AGT relation as Hubbard - Stratanovich duality which works for all values of $\beta$. In the appendix we summarize the main facts about geometry of Hilbert schemes of points and the instanton moduli spaces.

\section{AGT and Hubbard - Stratanovich duality \label{agthb}}
For simplicity, we consider here only the 4-point spherical case, i.e. the main object that will be considered here is the 4-point function on a sphere:
\be
\label{ConfBlock}
B(\Lambda)= \langle V_{\alpha_1}(0) V_{\alpha_{2}}(\Lambda) V_{\alpha_3}(1) V_{\alpha_4}(\infty) \rangle
\ee
where $V_{\alpha_i}(z)$ are the primary fields. The conformal dimensions and the central charge are parameterized as usual:
\be
\Delta_i=\alpha_{i}(Q-\alpha_i), \ \ \ c= 1 + 6 Q^2 \ \ \ Q=b+\frac{1}{b}
\ee
The proof suggested in \cite{MMS1}-\cite{MMS} consists of four steps:
\begin{itemize}
\item Using the Dostenko-Fateev integral representation rewrite the conformal block (\ref{ConfBlock})  in the following form:
\be
\label{ivar}
B(\Lambda) = \int d \mu(x)  \int d\mu(y) \, {\mathbb{{I}}}^2 (\Lambda)
\ee
where ${\mathbb{{I}}}^2(\Lambda)$ is the (tensor) square of the identity operator in the Fock space \footnote{This element might be considered as an operator:
$$
{\mathbb{{I}}} (1) : {\cal{F}} \longrightarrow {\cal{F}}
$$
With respect to the standard scalar product in the Fock space it, obviously, satisfies $\langle{\mathbb{{I}}} (\Lambda), f(x) \rangle=f(\Lambda y)$, such that at $\Lambda=1$ it is an identity operator.
}:
$$
{\mathbb{{I}}} (\Lambda)= \prod\limits_{i=1}^{n_-}\prod\limits_{j=1}^{n_+} (1-\Lambda x_i y_j)^{ \beta}, \ \ \ \textrm{and}  \ \ {\mathbb{{I}}}^2 (\Lambda)= \prod\limits_{i=1}^{n_-}\prod\limits_{j=1}^{n_+} (1-\Lambda x_i y_j)^{ 2 \beta}
$$
where the corresponding measures are given by the integrals:
$$
d \mu (x)= \prod\limits_{1 \leq i<j \leq n_+}(x_i-x_j)^{2 \beta} \, \prod\limits_{i=1}^{n_{+}} x_i^{2 \alpha_1 b} (1-x_i)^{2 \alpha_3 b} (1-\Lambda x_i)^{2 \alpha_2 b}
$$
and
$$
d \mu (y)= \prod\limits_{1 \leq i<j \leq n_{-}}(y_i-y_j)^{2 \beta} \, \prod\limits_{i=1}^{n_{-}} y_i^{2 \alpha_1 b} (1-y_i)^{2 \alpha_2 b} (1-\Lambda y_i)^{2 \alpha_3 b}
$$

where parameters $n_{+}$ and $n_{-}$ are the discrete parameters corresponding to the number of the screening operators in DF formalism.
\item
Use some orthonormal basis $K_{Y}(x)$ in the Fock space to represent the identity operator in the form:
\be
\label{Expansion}
{\mathbb{{I}}}^2(\Lambda)   = \sum\limits_{Y} \, \Lambda^{|Y|} K_{Y}(x) K^{\ast}_{Y}(y)
\ee
In the $SU(N)$ -case, the Fock space of corresponding conformal field theory is ${\cal{F}}^{\otimes N}$ where ${\cal{F}}$ is the fock space of free bosons. Thus, for $SU(2)$ -case that we consider here, the expansion in (\ref{Expansion}) runs over bipartitions $Y=(\lambda,\mu)$ and $|Y|=|\lambda|+|\mu|$. After this expansion, the conformal block takes the form
\be
\label{intr}
B(\Lambda)=\sum\limits_{Y}\, \Lambda^{|Y|}  \int d \mu(x) K_Y(x)  \int d \mu(y) K^{\ast}_{Y}(y)
\ee
The switch from the double integral over $x$ and $y$ to the double sum over bipartitions $Y=(\lambda,\mu)$ looks like a typical Hubbard-Stratanovich duality - thus the name of the entire approach.
\item Note that the remaining Selberg integrals are actually rational functions of paraments $n_+$ and $n_-$ (in more complicated examples of higher genus curves they are equally well defined functions on a Riemann surfaces which are expressible in corresponding theta functions)- and thus can be analytically continued to non-integer values of $n_{\pm}$.
\item Finally, after the standard switch of the variables, we can identify these rational functions with the coefficients of the Nekrasov partition function
$$
\int d \mu(x) K_{\lambda, \mu}(x)  \int d \mu(y) K^{\ast}_{\lambda, \mu}(y)=N_{\lambda, \mu}
$$
such that the conformal block takes the form of a sum over partitions and coincide with the Nekrasov function:
\be
B(\Lambda)=\sum\limits_{\lambda, \mu} \Lambda^{|\lambda|+|\mu|}  N_{\lambda,\mu}=Z^{Nek}(\Lambda)
\ee
\end{itemize}
\section{The choice of the basis $K_{Y}$ and problem with $\beta\neq 1$ \label{basch}}
The integrals in (\ref{intr}), of course, depend on the choice of a basis $K_{Y}$ for the intermediate states, thus the choice of these polynomials becomes a crucial point of the whole process. First of all, the polynomials $K_{Y}$ must form a basis, such that we would have some sort of Cauchy completeness identity for them. Second,  they should give rise to some reasonable Selberg integrals, i.e. to reproduce the coefficients of Nekrasov function the integrals must be completely factorizable  to  linear multiples.

In \cite{MMS,MMSS} this was achieved only for $c=1$ -  in this case the choice of the basis was the most naive: formed by a pair of two Schur functions $K_{\lambda,\mu}(x)= s_{\lambda}(x) s_{\mu}(x)$. Indeed for $c=1$ (what corresponds to $\beta=1$) we have:
\be
\label{beone}
{\mathbb{{I}}}^2 (\Lambda)= \prod\limits_{i=1}^{n_-}\prod\limits_{j=1}^{n_+} (1-\Lambda x_i y_j)^{ 2 }=\sum\limits_{\lambda_1, \lambda_2} \,\Lambda^{|\lambda|+|\mu|} s_{\lambda}( x ) s_{\mu} (x)  s_{\lambda}( y )   s_{\mu} (y)
\ee
As was shown in \cite{MMS}, in the  case  $\beta=1$ the integrals are of Selberg type, and they are indeed equal to the coefficients of Nekrasov function:
\be
\left. \int d \mu(x) s_{\lambda}( x ) s_{\mu} (x)\, \int d \mu(y) s_{\lambda}( y ) s_{\mu} (y)\right|_{\beta=1} =\left.N_{\lambda \mu}\,  \right|_{\epsilon_1+\epsilon_2=0}
\ee
Moreover, this continues to work after the $q$-deformation - to Macdonald polynomials and Jackson $q$-integrals which provides a proof \cite{MMSS} of simplest AGT relation for $5d$-theories at $c=1$.

However, the $\beta$-deformation to $c\neq 1$ breaks the agreement, both in $4d$ and $5d$. In fact, once the Jack polynomials appeared, it is clear that many things will be consistent with the $\beta$ - deformation  - and indeed they are. In $4d$, which we concentrate on in what follows, one can  deform (\ref{beone}) to:
\be
{\mathbb{{I}}}^{2 } (\Lambda)= \prod\limits_{i=1}^{n_-}\prod\limits_{j=1}^{n_+} (1-\Lambda x_i y_j)^{ 2 \beta}=\sum\limits_{\lambda, \mu} \,\Lambda^{|\lambda|+|\mu|} j_{\lambda}( x ) j_{\mu} (x)  j_{\lambda}( y )   j_{\mu} (y)
\ee
where the Schur functions $s_{\lambda}$ are now substituted by the Jack polynomials $j_{\lambda}$. Moreover, the integrals are still of Selberg type - \textbf{but not all of them are factorized to linear multiples as in $\beta=1$ case.} Worst of all is that \textbf{they do not coincide with the coefficients of Nekrasov functions}. Thus, the expansion in Jack polynomials gives $\Lambda$-expansion of the conformal block in bipartitions which, however, does not coincide with the expansion of Nekrasov partition function in factorized coefficients $N_{\lambda, \mu}$ (the total sum over partitions with fixed $|\lambda|+|\mu|$, of course,  gives the correct $\Lambda^{|\lambda|+|\mu|}$  - coefficient of $Z^{Nek}(\Lambda)$, but the individual terms do not coincide with $N_{\lambda,\mu}$ ). Furthermore, the individual Nekrasov coefficients $N_{\lambda,\mu}$ for $\beta\neq 1$ possess additional poles, which cancel in the sum over partitions and are spurious from the point of view of conformal block. All this left the situation with the proof of AGT for $\beta\neq 1$ unsatisfactory.

Clearly, what is needed, is some other choice of the basis $K_{Y}$, more adequate for description of Nekrasov coefficients. Conceptually, such a basis is provided by Nakajima construction, and technically its main difference from the above consideration is that the relevant functions $K_{Y}$ are no longer split into pairs of orthogonal polynomials:  they are new polynomials depending at once on the pair of partitions (or $N$ -partitions in the case of $SU(N)/W_{N}$ gauge/conformal field theory)  -  we call them generalized Jack polynomials $J_{\lambda \mu}(x) \neq j_{\lambda}(x) j_{\mu}(x) $. Instead of splitting, they decompose into a combination of ordinary Jack bilinears, but with coefficients which \textbf{depend on the Coulomb parameter $a$ (!)} - and this extra $a$-dependence (disappearing at $\beta=1$) is the reason why such a decomposition of modified DF integral was overlooked in the previous papers \cite{MMS,MMSS}.

As we demonstrate below, this approach is indeed successful: generalized Jack polynomials, extracted from the equivariant cohomologies of the instanton moduli spaces ${\cal{M}}_{n,r}$:

(i) provide a basis in the Fock space and give a relevant decomposition of identity ${\mathbb{{I}}}^2 (\Lambda)= \prod\limits_{i=1}^{n_-}\prod\limits_{j=1}^{n_+} (1-\Lambda x_i y_j)^{ 2 \beta}$

(ii) integrate to rational functions, which can be decomposed to linear multiples,
and can be easily continued to arbitrary values of $n_{\pm}$.

(iii) reproduce Nekrasov functions with all their spurious poles at $\beta\neq 1$.

This consideration also provides a clear link between the Dotsenko-Fateev and ABBFLT approaches to the proof of AGT and implies numerous straightforward generalizations in all possible directions (to $SU(N)$ -case, to $5d$ $q$-deformation and to $n$-point functions).

\section{Example at the first level \label{levone}}
In order to explain what happens when we switch from $j_{\lambda}$ to the generalized Jack polynomials $J_{\lambda,\mu}$ it is best to consider the simplest example. For this purpose we remind that the problem with the equality\footnote{To avoid overloading this paper with lengthy formulas, unneeded for the current purposes we refer for notation and further details for this example to original paper \cite{MMS} and \cite{MMSS}. In this concrete case - to eqs. (12)-(16) of \cite{MMSS}.  In short, the time variables are related to the integration variables in (\ref{ivar}) by the Miwa transform $p_{k}=\sum_{i=1}^{n_+} x_{i} $, $\bar{p}_{k}=\sum_{i=1}^{n_-} y_{i} $. The  variables $v_{\pm}$ and $u_{\pm}$ are linear combination of masses, integrals over $x$ and $y$ denoted by $\langle\rangle_{\pm}$ and the crucial parameter $a$ is expressed through the quantities of integration $n_{\pm}$. }:
\be
\sum\limits_{\lambda \mu} \, N_{\lambda, \mu} = \sum\limits_{\lambda \mu} \, \Big\langle j_{\lambda}(-p_k -\frac{v_+}{\beta}) j_{\mu}(p_k) \Big\rangle_{+}  \Big\langle j_{\lambda}(-\bar{p}_k ) j_{\mu}(-\bar{p}_k-\frac{v_-}{\beta}) \Big\rangle_{-}
\ee
For $\beta\neq 1$ this problem appears already at the level one, $|\lambda| +|\mu|=1$, and even for the vanishing masses, when $v_+=v_-=0$:
since $j_{[]}=1$ and $j_{[1]}=p_1$,  we have:
$$
N_{[1],[]}+N_{[],[1]}=\langle j_{[1]}  \rangle_{+} \langle j_{[1]}  \rangle_{-}+\langle j_{[1]}  \rangle_{+} \langle j_{[1]}  \rangle_{-}=- \langle p_1 \rangle_{+} \langle p_1 \rangle_{-}-\langle p_1 \rangle_{+} \langle p_1 \rangle_{-}
$$
(i.e. the sum of two equal terms) after the substitution of the explicit expressions for Nekrasov functions on the left side and  Selberg integrals on the right we obtain an identity:
$$
\dfrac{1}{2 a (2 a -\epsilon)} + \dfrac{1}{2 a (2 a +\epsilon)}=\dfrac{1}{(2a-\epsilon) (2a+\epsilon)}+\dfrac{1}{(2a-\epsilon) (2a+\epsilon)}
$$
with $\epsilon=\epsilon_1+\epsilon_2$ in Nekrasov's notations. This identity is of course correct, but the individual terms on the left and right side do not match, i.e. individual Nekrasov coefficients $N_{\lambda, \mu}$ are not reproduced by averages of Jack polynomials. However, at $\epsilon=0$ corresponding to $\beta=1$ the above identity becomes termwise.

Introduction of masses makes this discrepancy  more profound, in this case we have:
$$
N_{[1],[]}+N_{[],[1]}=\Big\langle j_{[1]}(-p_k-\frac{v_+}{\beta})  \Big\rangle_{+} \Big\langle j_{[1]}(\bar{p}_k)  \Big\rangle_{-}+\Big\langle j_{[1]}(p_k)  \Big\rangle_{+} \Big\langle j_{[1]}(-\bar{p}_k-\frac{v_{-}}{\beta})  \Big\rangle_{-}
$$
or, substituting  the explicit expressions for Jack polynomials:
\be
\label{nnn}
N_{[1],[]}+N_{[],[1]}=-(\langle p_1 \rangle_{+} +v_{+}) \langle \bar{p}_1\rangle_{-}- \langle p_1\rangle_{+} (\langle \bar{p}_1 \rangle_{-} +v_{-})
\ee
Again, after substitution of the explicit expressions for Nekrasov functions on the left and Jack correlators on the right side we obtain the following identity:
\be
\label{dn}
\begin{array}{c}
\dfrac{\prod\limits_{i=1}^{4}(a+m_i)}{ 2a (2a +\epsilon)} + \dfrac{\prod\limits_{i=1}^{4}(a-m_i)}{ 2a (2a -\epsilon)} =\\
\\
\left( \dfrac{(a-m_1)(a-m_2)}{\underline{2a-\epsilon}} + (m_1+m_2) \right)  \dfrac{(a+m_3)(a+m_4)}{2a+\epsilon} + \dfrac{(a-m_1)(a-m_2)}{2a-\epsilon} \left( \dfrac{(a+m_3)(a+m_4)}{\underline{2a+\epsilon}} - (m_3+m_4) \right)
\end{array}
\ee
Exactly as in the case without masses, the sum of two terms on the right is equal to the sum of two terms on the left, but individually terms do not coincide.
Note, that particular Nekrasov functions at the left side have extra (spurious) pole (at $a=0$), which is not present neither in the entire sum, nor in the particular Selberg integrals at the right side. In general these are poles beyond the zeroes of Kac  determinant, and their probable \textit{raison d'etre} of $U(1)$ factor in the AGT relations well emphasized in the approach \cite{Alba1}.

Clearly, to cure the problem one needs to replace both underlined denominators at the right side of (\ref{dn}) by $2a$. This, however, means that at the right side of (\ref{nnn}) one should somehow get:
\be
\label{rexp}
-\Big(\frac{2 a -\epsilon}{ 2 a } \langle p_1 \rangle_{+} +v_{+}\Big) \langle \bar{p}_1\rangle_{-}- \langle p_1\rangle_{+} \Big(\frac{2 a +\epsilon}{ 2 a } \langle \bar{p}_1 \rangle_{-} +v_{-}\Big)
\ee
Of course, this does not change the full answer: what adds to one term is subtracted in another, however the splitting of the answer in two terms changed - and in explicitly $a$-dependent way: we move $\frac{\epsilon}{2 a} \langle p_1 \rangle_{+}\langle \bar{p}_1 \rangle_{-}$ from one term to another.

The main claim is that this is exactly what happens, when one substitutes the expansion in $j_{\lambda}$ by that in generalized Jack polynomials
$J_{\lambda, \mu}$:
\be
\sum\limits_{\lambda \mu} \, N_{\lambda, \mu} = \sum\limits_{\lambda \mu} \, \Big\langle J_{\lambda,\mu}(-p_k -\frac{v_+}{\beta}| p_k)  \Big\rangle_{+}  \Big\langle \bar{J}_{\lambda, \mu}(\bar{p}_k | -\bar{p}_k-\frac{v_-}{\beta}) \Big\rangle_{-}
\ee
such that this identity becomes termwise (!):
\be
 N_{\lambda \mu}=\Big\langle J_{\lambda,\mu}(-p_k -\frac{v_+}{\beta}| p_k)  \Big\rangle_{+}  \Big\langle \bar{J}_{\lambda, \mu}(\bar{p}_k | -\bar{p}_k-\frac{v_-}{\beta}) \Big\rangle_{-}
\ee
The new polynomials $J_{\lambda, \mu}$, which substitute $j_{\lambda} j_{\mu}$, naturally depend on two set of time-variables, and this is actually the key point. Keeping this in mind, it is easy to write down a set of polynomials (and its dual):
\be
\label{feg}
\begin{array}{ll}
 J_{[1],[]}(p_k|\bar{p}_k)=p_1-\dfrac{\epsilon}{2 a} \bar{p}_{1},  \ \ \ &  J_{[1],[]}^{\ast}(p_k|\bar{p}_{k})=p_1, \\
\\
 J_{[],[1]}(p_k|\bar{p}_k)= \bar{p}_1, \ \ \ &  J_{[],[1]}^{\ast}(p_k|\bar{p}_{k})=\bar{p}_1+\dfrac{\epsilon}{2 a} p_1
\end{array}
\ee
Such that the decomposition of the identity element in the Fock space (Cauchy completeness identity) takes the form:
\be
{\mathbb{{I}}}^2 (\Lambda)= \prod\limits_{i=1}^{n_-}\prod\limits_{j=1}^{n_+} (1-\Lambda x_i y_j)^{ 2 \beta}=\sum\limits_{\lambda \mu} \, \Lambda^{|\lambda|+|\mu|} j_{\lambda}(p_k) j_{\lambda}(\bar{p}_k) j_{\mu}(p_k) j_{\mu}(\bar{p}_k) = \sum\limits_{\lambda \mu} \,\Lambda^{|\lambda|+|\mu|} J_{\lambda ,\mu}(p_k|\bar{p}_k) \,J^{\ast}_{\lambda, \mu}(p_k|\bar{p}_k)
\ee
and reproduces (\ref{rexp}) at the first level:
\be
(\ref{rexp})= \Big\langle J^{\ast}_{[1],[]}(-p_k-\frac{v_+}{\beta}|p_k) \Big\rangle_{+} \Big\langle J_{[1],[]}(\bar{p}_k|-\bar{p}_k-\frac{v_-}{\beta}) \Big\rangle_{-} +  \Big\langle J^{\ast}_{[],[1]}(-p_k-\frac{v_+}{\beta}|p_k) \Big\rangle_{+} \Big\langle J_{[],[1]}(\bar{p}_k|-\bar{p}_k-\frac{v_-}{\beta}) \Big\rangle_{-}
\ee
This example exhaustively explains what happens when we switch to expansion in generalized Jack polynomials.

What is important, however, these polynomials are not only adjusted to AGT relation, they have an \textit{a priori } definition in terms of the equivariant cohomologies of the instanton moduli spaces, which will be given in section \ref{apb}. Since  this is a "natural" definition, it can be considered as providing a proof of the AGT relation: \textbf{conformal block is $DF$ average of the $\beta$-squared "Vandermonde determinant", which being expanded in the generalized Jack polynomials,   decomposes into sum of the Nekrasov functions.}

\section{Special polynomials as eigenfunctions \label{hamilt}}
The most practical way to define Schur polynomials
and their generalizations is as eigenfunctions
of certain differential or difference operators.
If polynomials are expressed through the time-variables $p_k$,
they are known as cut-and-join operators,
if a Miwa transform is performed,
$p_k = \Tr_{N\times N} X^k = \sum_\alpha x_\alpha^k$,
then the operators are given by  Hamiltonians of Calogero-Moser-Sutherland (CMS)
integrable systems.
Schur polynomials {\it per se} satisfy
\be
\hat H s_\lambda = \varphi_\lambda s_\lambda
\label{Seq}
\ee
with
\be
\begin{array}{l}
\hat H = \dfrac{1}{2}  :\tr \left(X\dfrac{\partial}{\partial X^{tr}}\right)^2: =\dfrac{1}{2}
\sum\limits_{i=1}^N x_i^2\dfrac{\partial^2}{\partial x_i^2}
+ \dfrac{1}{2}\sum\limits_{i\neq j}\dfrac{x_i x_j}{x_i-x_j}\left(\dfrac{\partial}{\partial x_i}-
\dfrac{\partial}{\partial x_j}\right)=
\\
\\ \dfrac{1}{2} \sum\limits_{n,m} \left(n m p_{n+m}\dfrac{\partial^2}{\partial p_n\partial p_m}
+ (n+m)p_n p_m\dfrac{\partial}{\partial p_{n+m}}\right)
\end{array}
\label{Sop}
\ee
and dependence on the Young diagram $\lambda$ is controlled by the
eigenvalue
\be
\varphi_\lambda = \sum_{(i,j)\in\lambda} (i-j)
\label{Sev}
\ee
The equality between  operators in $p$ and $x$-variables takes place,
when they act on functions of $X$, i.e. on the $N$-dimensional subspace
in the space of the time-variables.
In fact, Schur functions are the characters of the linear group $GL(N)$,
they are common eigenfunctions of the infinite set of cut-and-join
operators $\hat H_\mu$, also labeled by Young diagrams $\mu$
and the eigenvalues $\varphi_\lambda(\mu)$ are the characters of the
symmetric (permutation) group $S_N$, see \cite{MMN} for details.
However, for technical purpose of building up the polynomials $s_\lambda(p)$
just (\ref{Seq}),(\ref{Sop}) and (\ref{Sev}) are enough.
Jack polynomials $j_\lambda(p)$ are characterized by a $\beta$-deformed
version of (\ref{Seq})-(\ref{Sev}):
\be
\label{dh}
\hat H^{(\beta)} j_\lambda = \varphi_\lambda^{(\beta)} j_\lambda
\ee
with
\be
\hat H^{(\beta)} = \dfrac{1}{2} \sum\limits_{n,m=1}^{\infty}\,\Big(n m p_{n+m}
\dfrac{\partial^2}{ \partial p_n \partial p_m}+
 \beta (n+m) p_n p_m \dfrac{\partial}{ \partial p_{n+m}}\Big)
 -\dfrac{1-\beta}{2} \sum\limits_{n=1}^{\infty} \, (n-1) n p_n \dfrac{\partial}{\partial p_n }
\label{Jop}
\ee
and
\be
\varphi_\lambda^{(\beta)} = \sum_{(i,j)\in\lambda} \Big((i-1)-(j-1)\beta\Big)
\ee
such that at $\beta=1$ formulae (\ref{dh})-(\ref{Jop}) degenerate to (\ref{Seq})-(\ref{Sev}) and Jack polynomials turn to Schur polynomials.
Further, $q$-deformation provides the set of Macdonald polynomials,
relevant for description of $5d$ AGT relations \cite{MMSS}.
As to generalized Jack polynomials $J_{\lambda,\bar\lambda}(p,\bar p)$ (in the $SU(2)$ -case)
needed for the proof of AGT relations,
they satisfy
\be
\boxed{
\hat{\cal H}^{(\beta)} J_{\lambda,\bar\lambda}
= \Phi_{\lambda,\bar\lambda}^{(\beta)}(u,\bar u) J_{\lambda,\bar\lambda}
}
\ee
with
\be
\begin{array}{r}
\hat{\cal H}^{(\beta)} = \dfrac{1}{2} \sum\limits_{n,m=1}^{\infty}\, \Big(\beta (n+m) p_n p_m  \dfrac{\partial}{ \partial p_{n+m}} + n m p_{n+m} \dfrac{\partial^2}{ \partial p_n \partial p_m}\Big) +\dfrac{1}{2} \sum\limits_{n=1}^{\infty} \, \Big(2 u+(\beta-1)(n-1)\Big) n p_n \dfrac{\partial}{\partial p_n }+\\
\\
\dfrac{1}{2} \sum\limits_{n,m=1}^{\infty}\Big(\beta (n+m) \bar{p}_n \bar{p}_m
\dfrac{\partial}{ \partial \bar{p}_{n+m}} +
n m \bar{p}_{n+m} \dfrac{\partial^2}{ \partial \bar{p}_n \partial \bar{p}_m} \Big)+\dfrac{1}{2}
\sum\limits_{n=1}^{\infty} \Big(2 \bar u +(\beta-1)(n-1)\Big) n \bar{p}_n
\dfrac{\partial}{\partial \bar{p}_n } +\\
\\
\underline{
(1-\beta)\sum\limits_{n=1}^{\infty} n^2 \bar{p}_{n} \dfrac{\partial}{\partial p_n }}
\end{array}
\label{JJop}
\ee
and
\be
\Phi_{\lambda,\bar\lambda}^{(\beta)} = \sum_{(i,j)\in\lambda} \Big( u +(i-1)-(j-1)\beta\Big) + \sum_{(i,j)\in \bar \lambda} \Big( \bar u +(i-1)-(j-1)\beta\Big)
\ee
Note that the last (underlined) term makes
this $\hat{\cal H}$ not Hermitian, therefore the conjugate
Jack polynomials $J^*_{\lambda,\bar\lambda}$ satisfy a slightly different
equation:
\be
\boxed{
\bar{\cal H}^{(\beta)}J^*_{\lambda,\bar\lambda}=
\Phi_{\bar\lambda,\lambda,}^{(\beta)}(\bar u, u) J^*_{\lambda,\bar\lambda}
}
\ee
Another feature of crucial importance is that the Hamiltonian
-- and thus its eigenfunctions and eigenvalues -- depend on
additional set of "spectral" parameters $u_1,\ldots,u_N$.
In the case of $N=2$ we identify $ u =-\bar u$ with $a$.
For example,
$J_{[],[1]} = \bar p_1$ is the obvious eigenfunction with eigenvalue  $\bar  u= - a$.
However, this is not true about $ p_1$ --
eigenfunction is only the linear combination:
$J_{[1],[]}= p_1+\frac{1-\beta}{2a} \bar p_1$ with eigenvalue $ u =a$
-- exactly the one that we needed in (\ref{feg}).

This description is sufficient to build up all the
$
J_{\lambda,\bar\lambda}$ and check some of their properties.
Generalizations to other groups, to $5d$ and to other conformal blocks
looks straightforward but
requires more detailed analysis.
Also open is the question about the $\beta$-deformation of entire center of
the symmetric group algebra, which would provide the full set of
deformed cut-and-join operators {\it a la} \cite{MMN}.
This problem is deeply related to $3d$ AGT relations in the spirit of
\cite{3dAGT}, in particular to genus expansion of superpolynomials,
generalizing the results of \cite{DMMSS}-\cite{Anton2}.
In the remaining sections we comment briefly on relation of the
operator $\hat{\cal H}^{(\beta)}$ to the theory of equivariant cohomologies of the
instanton moduli spaces, which generalizes a similar relation
of ordinary Schur and Jack polynomials to equivariant cohomologies
of Hilbert schemes,   and on the Selberg-Kadell \cite{Selb}-\cite{Kadel2} factorization properties
of the integrals of generalized Jack polynomials.
In fact this presentation is immediately applicable to arbitrary $GL(N)$ and thus to the $W_N$ conformal blocks.
\section{Selberg averages of generalized Jack polynomials \label{saver}}
An important family of integrals, appearing in DF representation of $GL(N)$ conformal blocks are the so called $GL(N)$-Selberg integrals. In the simplest $GL(2)$-case, which we concentrate on in this paper,  the Selberg integral is formulated as the following average:
\be
\label{sa}
\Big\langle f({x_1,...,x_n})\Big\rangle=\dfrac{1}{Z}\, \int_{0}^{1} \, \prod\limits_{i<j}(x_i-x_j)^{2 \beta} \, \prod\limits_{i=1}^n x_i^u (1-x_i)^v dx_i  f({x_1,...,x_n})
\ee
for some symmetric polynomial $f({x_1,...,x_n})$ with
$$
Z=\int_{0}^{1} \, \prod\limits_{i<j}(x_i-x_j)^{2 \beta} \, \prod\limits_{i=1}^n x_i^u (1-x_i)^v dx_i
$$
here $\beta$ is usually assumed to be integer. The non-integer values of $\beta$ arise as the analytical continuation of this function.

In the case of some special $f({x_1,...,x_n})$, this average takes closed, \textit{completely factorized} form. For example, in the case when the integrand is given by a Jack polynomial $f(x_i)=j_{\lambda}(x_i)$ Kadell \cite{Kadel1,Kadel2} proved the following formula:
\be
\label{Kad1}
\Big\langle j_\lambda({x_1,...,x_n})\Big\rangle=\dfrac{\tau_{\lambda}( u+n\beta+1-\beta )\, \tau_\lambda( n\beta)}{\tau_{\lambda}(u+v+2 n \beta+2-2\beta)}
\ee
where the function:
\be
\tau_{\lambda}(u)=\dfrac{1}{\beta^{|\lambda|}}\prod\limits_{(i,j) \in \lambda} \, \Big(u+(i-1) -(j-1)\beta\,\Big)
\ee
has a geometrical meaning of the Euler class of the  tautological bundle over Hilbert schemes (see sections \ref{apa}-\ref{apb}).

Natural object arising in DF representation of $SU(2)$ conformal block, is the Selberg average of two Jack polynomials. In \cite{MMS} was found the following formula for such an average:
\be
\label{SHf}
\begin{array}{l}
\Big< j_\lambda(p_k + w) j_\mu(p_k) \Big>\ = \ \dfrac{1}{{\rm Norm_{\beta}(u,v,n)}} \dfrac{\tau_{\lambda}(v+n\beta+1-\beta) \tau_{\mu}(u+n\beta+1-\beta)}{\tau_{\lambda}(n \beta)
\tau_{\mu}(u+v+n\beta+2-2\beta)} \times \emph{} \\
\\
\emph{} \times
\dfrac{\prod\limits_{i<j}^{n} \Big( \lambda_i - \lambda_j + (j-i)\beta\Big)_{\beta} \prod\limits_{i<j}^{n} \Big( \mu_i - \mu_j + (j-i)\beta\Big)_{\beta} }
{\prod\limits_{i,j}^{n} \Big( \ u+v+2\beta n + 2+\lambda_i+\mu_j - (1+i+j)\beta \ \Big)_{\beta}}
\end{array}
\ee
where the shift is $w = (v + 1 - \beta)/\beta$ and $\lambda,\mu$-independent normalization constant ${\rm Norm}_{\beta}(u,v,N)$ is defined from the condition
$\Big< 1 \Big> = 1$. The  Pochhammer symbol $(x)_{\beta}$ is defined as
\begin{align}
(x)_{\beta} = \dfrac{\Gamma(x+\beta)}{\Gamma(x)} = x(x+1)\ldots(x+\beta-1)
\end{align}
Despite the fact that  this formula gives some expression for the average of two Jack polynomials, completely factorized to linear multiples, it can not be applied to the AGT relation for a number of reasons. First, the shift appearing in DF integrals does not coincide with $w$. Second, (\ref{SHf}) expressed in terms of Nekrasov variables this average coincide with the coefficient $N_{\lambda,\mu}$ of Nekrasov partition function only for $\beta=1$ \cite{MMS}.

The main result of this paper is the following generalization of Kadell integrals: we discovered that the $GL(2)$ Selberg averages of corresponding
$GL(2)$ generalized Jack polynomials have a closed completely factorized form. Moreover, the answer is much simpler then the Kadell's formula:
under the following relation between $a$ and $n$ known in AGT conjecture:
\be
\label{un}
a=-\beta n -\frac{1}{2} ( u+v+1-\beta)
\ee
we have\footnote{ This formulae are checked on a computer for $|\lambda|+|\mu| \leq 6$.   }:
\be
\label{Av1}
\begin{array}{|c|}
\hline\\
\Big\langle  J_{\lambda,\mu}(a,-p_k-v/\beta,p_k) \Big\rangle=(-1)^{|\lambda|+|\mu|} \tau_{\lambda}(-v-\beta n) \tau_{ \lambda}(-u-v-\beta n -1+\beta) \tau_{\mu}(\beta n)
\tau_{\mu }(u+\beta n+1-\beta)
\\
\\
\hline
\end{array}
\ee
Analogous formula for dual basis, can be obtained from above formula applying  (\ref{drd}):
\be
\label{Av2}
\begin{array}{|c|}
\hline\\
\Big\langle\,   J^{\ast}_{\lambda,\mu}(a,p_k,-p_k-v/\beta)   \, \Big\rangle=(-1)^{|\lambda|+|\mu|} \tau_{\lambda}(\beta n) \tau_{ \lambda}(u+\beta n+1-\beta) \tau_{\mu}(-v-\beta n)
\tau_{\mu}(-u-v-\beta n-1+\beta)
\\
\\
\hline
\end{array}
\ee
Note, that in the normalization of Jack and generalized Jack polynomials accepted in this paper, (see section \ref{apa}) we have:
\be
J_{[],\lambda}(a)=(-1)^{|\lambda|} \tau_\lambda({-2a+1-\beta}) j_{\lambda}
\ee
such that the first formula of Kadell (\ref{Kad1}) is a simple corollary of (\ref{Av1}).

In the next section we show that these formulae applied to DF representaton of conformal block lead directly to instanton part of Nekrasov partition function, and reproduce all individual coefficients for arbitrary $\beta$. The generalization of these formulae to the case of $SU(N)$ - Selberg integrals \cite{ZM} and rank $N$ generalized Jack polynomials looks straightforward, but remains to be completed elsewhere.

\section{Proving AGT relation  \label{sproof}}
Let us consider the simplest CFT four point function on a sphere:
$$
B(q)=\langle V_{\alpha_1}(0) V_{\alpha_2}(q) V_{\alpha_3}(1) V_{\alpha_4}(\infty)  \rangle
$$
Using the Dotsenko-Fateev representation \cite{DF1,DF2}, we can express this function as integrated free field correlator:
\be
B(q)=\langle :e^{\alpha_1 \phi(0)}: :e^{\alpha_2 \phi(q)}: :e^{\alpha_3 \phi(1)}: :e^{\alpha_4 \phi(\infty)}:  \Big(\int_{0}^{q} :e^{b \phi(z) }: dz \Big)^{n_+}
\Big(\int_{1}^{\infty} :e^{b \phi(z) }: dz \Big)^{n_-}   \rangle
\ee
where $b^2=\beta$. Using free field identity:
$$
\langle :e^{\alpha_1 \phi(z_1)}:...:e^{\alpha_n \phi(z_n)}: \rangle=\prod\limits_{i<j} \,(z_i-z_j)^{2 \alpha_i \alpha_j}
$$
after a simple change of variables we obtain the following representation of the conformal block  (see section 2 in  \cite{Ito} for details):
\be
B(\Lambda)=\Big\langle \, \Big\langle \,\prod\limits_{i=1}^{n_-} (1-\Lambda x_i)^{v_-} \prod\limits_{i=1}^{n_+} (1-\Lambda y_i)^{v_+}  \prod\limits_{i=1}^{n_-}\prod\limits_{j=1}^{n_+} (1-\Lambda x_i y_j)^{2 \beta}\, \Big\rangle_{+} \, \Big\rangle_{-}
\ee
where $\langle\rangle_{\pm}$ are two independent Selberg averages (\ref{sa}) with $u = u_\pm$ and $v=v_\pm$. These parameters
$v_\pm$, $u_\pm$ and $n_\pm$ are obviously related to the conformal dimensions $\alpha_i$ of the conformal block. After some simple algebra we rewrite it in the exponential form as:
\be
B(\Lambda)=\Big\langle \, \Big\langle \,  \exp\left ( \beta \sum\limits_{k=1}^{\infty}\, \frac{\Lambda^k}{k} \Big( p_{k}  (-q_k- \frac{v_-}{\beta})  + q_k (-p_k-\frac{v_+}{\beta})  \Big) \right)   \, \Big\rangle_{+} \, \Big\rangle_{-}
\ee
Now, there are two natural ways to expand it in polynomials: the first is the expansion in Jack polynomials which utilizes the Cauchy identity (\ref{Jcauchy}):
\be
\label{exp1}
\begin{array}{l}
B(\Lambda)=\Big\langle \, \Big\langle \,  \sum\limits_{\lambda, \mu} \, \Lambda^{|\lambda|+|\mu|} \dfrac{j_{\lambda}( p_k )  j_\lambda (-q_k- \frac{v_-}{\beta})}{ e_{\lambda, \lambda }(0)} \dfrac{j_{\mu}( q_k )  j_\mu (-p_k- \frac{v_+}{\beta})}{ e_{\mu, \mu }(0)}    \, \Big\rangle_{+} \, \Big\rangle_{-} =\\
\\
= \sum\limits_{\lambda, \mu} \, \dfrac{\Lambda^{|\lambda|+|\mu|}}{  e_{\lambda, \lambda }(0) \, e_{\mu, \mu }(0)}  \, \Big\langle j_{\lambda}( p_k ) j_\mu (-p_k- \frac{v_+}{\beta}) \Big\rangle_{+} \Big\langle j_{\mu}( q_k  )   j_\lambda (-q_k- \frac{v_-}{\beta}) \Big\rangle_{-}
\end{array}
\ee
This expansion were studied in \cite{MMS,MMSS} -  the main problem is that it does not reproduce the expansion of conformal block in the form of Nekasov partition function for $\beta\neq 1$. The correlators of two Jack polynomials is factorizing to linear terms only for $\beta=1$:
\be
\label{lolo}
\left. \frac{ \Big\langle j_{\lambda}( p_k ) j_\mu (-p_k- \frac{v_+}{\beta}) \Big\rangle_{+} \Big\langle j_{\mu}( q_k  )   j_\lambda (-q_k- \frac{v_-}{\beta}) \Big\rangle_{-}}{  e_{\lambda, \lambda }(0) \, e_{\mu, \mu }(0) }\right|_{\beta=1} =\left. N_{\lambda,\mu}\right|_{\, \epsilon_1+\epsilon_2=0}
\ee
where $N_{\lambda, \mu}$ is the coefficient of Nekrasov partition function for $SU(2)$ theory, Thus at $\beta=1$ we have an identity:
\be
B(\Lambda)=\sum\limits_{\lambda, \mu} \, \Lambda^{|\lambda|+|\mu|} \, N_{\lambda, \mu} = Z^{Nek}(\Lambda)
\ee
For $\beta\neq1$ the situation is  different: the Selberg averages of two Jack polynomials  with a shift $v_{\pm}/\beta$ appearing in (\ref{lolo}) are not completely factorized. The correlators have different structure of poles
in the parameter $a$ (see discussion of poles problem in \cite{MMS} ) and the expansion  of conformal block (\ref{exp1}) does not reproduce the expansion of Nekrasov partition functions.

The solution to this problem is different choice of the basis: we can rewrite the integrand in the basis of the generalized Jack polynomials using (\ref{GJcauchy}):
\be
\label{trr}
 \exp\left ( \beta \sum\limits_{k=1}^{\infty}\, \frac{\Lambda^k}{k} \Big( p_{k}  (-q_k- \frac{v_-}{\beta})  + q_k (-p_k-\frac{v_+}{\beta})  \Big) \right) =\sum\limits_{\mu,\nu}\,\Lambda^{|\mu|+|\nu|}\,\dfrac{J_{\mu,\nu} (a,-p_k-v_+/\beta,p_k)  J_{\mu,\nu}^{\ast} (a,q_k,-q_k-v_-/\beta)}{e_{\mu,\mu}(0) \,e_{\mu,\nu}(2 a)\, e_{\nu,\mu}(- 2 a) e_{\nu,\nu}(0)}
\ee
Note, that a new parameter $a$ appears on the right side of this decomposition which is absent on the left. Obviously, the whole expression here does not depend on $a$, and this parameter cancels  after summation. To make Selberg averages factorize, we need some wise choice of this parameter.
This choice is given explicitly by (\ref{un}). Thus for the conformal block we obtain:
\be
B(\Lambda)=\sum\limits_{\mu,\nu}\,\Lambda^{|\mu|+|\nu|}\,\dfrac{ \Big\langle J_{\mu,\nu} (a,-p_k-v_+/\beta,p_k)  \Big\rangle_{+} \Big\langle  J^{\ast}_{\mu,\nu} (a,q_k,-q_k-v_-/\beta)  \Big\rangle_-}{e_{\mu,\mu}(0) \,e_{\mu,\nu}(2 a)\, e_{\nu,\mu}(- 2 a) e_{\nu,\nu}(0)}
\ee
Due to the special choice of $a$, the averages factorize and are given explicitly by (\ref{Av1}) and (\ref{Av2}).
The final point here is to note that after usual AGT change of variables:
\be
\label{masses}
\begin{array}{l}
\beta=-\dfrac{\epsilon_1}{\epsilon_2}, \ \ n_+=\dfrac{a-\mu_2}{\epsilon_1}, \ \ n_-=\dfrac{-a-\mu_4}{\epsilon_1},
\\
\\
u_+=\dfrac{\mu_1-\mu_2-\epsilon_1-\epsilon_2}{\epsilon_2}, \ \ \ u_-=\dfrac{\mu_3-\mu_4-\epsilon_1-\epsilon_2}{\epsilon_2},
\\
\\
v_+=\dfrac{-\mu_1-\mu_2}{\epsilon_2}, \ \ \  v_-=\dfrac{-\mu_3-\mu_4}{\epsilon_2}
\end{array}
\ee
these averages exactly reproduce the coefficients of Nekrasov partitions function for arbitrary $\beta$:
\be
N_{\mu\nu}=\dfrac{ \Big\langle J_{\mu,\nu} (a,-p_k-v_+/\beta,p_k) \Big\rangle_{+} \Big\langle J^{\ast}_{\mu,\nu} (a,q_k,-q_k-v_-/\beta)  \Big\rangle_-}{e_{\mu,\mu}(0) \,e_{\mu,\nu}(2 a)\, e_{\nu,\mu}(- 2 a) e_{\nu,\nu}(0)},
\ee
and the identity between the four point conformal block and the Nekrasov partition function turns out termwise:
$$
B(\Lambda)=\sum\limits_{\mu,\nu}\,\Lambda^{|\mu|+|\nu|} N_{\mu\nu} = Z_{Nek}(\Lambda)
$$
At $\beta=1$ the generalized Jack polynomials degenerate to a product of two Schur polynomials:
\be
| J_{\mu,\nu}(p_k,q_k) \rangle \sim s_{\mu}(p_k) s_{\nu}(q_k)
\ee
and the above result reproduces the result of \cite{MMS}. For $\beta \neq 1$ this approach resolves the poles problem addressed in \cite{MMS}.
The appearance of new poles happens already at the level of Cauchy identity (\ref{trr}), in which both sides (explicitly on the right side and unexplicitly on the left side) are independent of the parameter $a$.

\section{Appendix A: Jack polynomials and Hilbert schemes \label{apa}}
\subsection{}
Here we give a short outline of relation between the Jack polynomials and the classes of fixed points in the equivariant cohomologies of Hilbert schemes. The details of this construction can be found in \cite{Nak1,Nak4,OP,LQW,Vass}. The Hilbert schemes of $n$ points in the plane ${\mathbb{C}}^2$  is defined as the space of polynomial ideals in two variables $x$ and $y$ which have codimension~$n$:
$$
\textrm{Hilb}_{n}= \{ {\cal{J}}\subset {\mathbb{C}}[x,y]:  \dim \Big( {\mathbb{C}}[x,y]/{\cal{J}} \Big)=n \}
$$
Alternatively, we can think about $\textrm{Hilb}_{n}$ as moduli space of $n$ points in ${\mathbb{C}}^2$.
Assume that the number of points is partitioned $n=n_1+...+n_k$. Then, we can consider a subset of $\textrm{Hilb}_{n}$ representing the union of $n_1$ points sitting at the same point,  $n_2$ all sitting at some other point in $ {\mathbb{C}}^2$ and so on. These subsets, labeled by partitions of $n$, give some cycles in cohomologies of Hilbert scheme. Moreover, these cycles form a basis of cohomologies such that:
$$
\dim H^{\bullet}(\textrm{Hilb}_{n})  =p(n)
$$
where $p(n)$ is the number of partitions of $n$. Thus, it is natural to consider the "composed" space $\coprod\limits_{n=1}^{\infty} \textrm{Hilb}_{n}$ such that its cohomologies can be naturally identified with the space of polynomials on infinite number of variables (i.e. boson Fock space):
\be
H^{\bullet}\Big(\coprod\limits_{n=1}^{\infty} \textrm{Hilb}_{n} \Big)= {\mathbb{C}}[p_1,p_2,...]\simeq {\cal{F}}
\ee
such that a partition $n=n_1+...+n_k$ corresponds to the element:
$$
p_{n_1}p_{n_2}...p_{n_k}\in {\mathbb{C}}[p_1,p_2,...]
$$
The \textit{cohomological} degree of this elements is defined by $\deg p_{k}=2(k-1)$, such that the element $p_{1}^n$ represents the unit
and $p_n$  - the top class in $H^{\bullet}(\textrm{Hilb}_{n})$. For example, the cohomologies of $\textrm{Hilb}_{4}$ are spanned by the following elements:
$$
H^{0}(\textrm{Hilb}_{4})=\textrm{Span}\{  p_1^4 \}, \ \ H^{2}(\textrm{Hilb}_{4})=\textrm{Span}\{  p_2 p_1^2 \}, \ \ H^{4}(\textrm{Hilb}_{4})=\textrm{Span}\{  p_3 p_1,\,  p_2^2 \}
, \ \ H^{6}(\textrm{Hilb}_{4})=\textrm{Span}\{  p_4  \}
$$

\subsection{}
Any element $\gamma  \in H^{k}(\textrm{Hilb}_{n})$ defines the operator of cup product:
$$
 \hat{\gamma} : H^{\bullet}(\textrm{Hilb}_{n}) \longrightarrow H^{\bullet+k}(\textrm{Hilb}_{n}), \ \  \hat{\gamma}(c) = \gamma \cup c
$$
The most important of these are the operators representing the characteristic classes of tautological bundle over the Hilbert scheme. The tautological bundle over $\textrm{Hilb}_{n}$ is rank$=n$ bundle with the fiber ${\mathbb{C}}[x,y]/{\cal{J}}$ over a point represented by the ideal ${\cal{J}}$.

Let $c_1$ be the first Chern class of this bundle. By definition, it has degree two, at the same time, the spaces $H^{0}(\textrm{Hilb}_{n})$ and $ H^{2}(\textrm{Hilb}_{n})$ are always one-dimensional and are spanned by $p_1^n$ and $p_2 p_1^{n-2}$ respectively.
Thus, up to a coefficient we should have $ \hat{c}_1( p_1^n) \sim p_2 p_1^{n-2}$, i.e.  it should contain the following term:
$$
\hat{c}_1\sim p_2 \dfrac{\partial^2}{\partial^2 {p_1}}+...
$$
where the dots stand for the higher terms.  The general formula for the first Chern class was found by Lehn \cite{Lehn1,Lehn2}:
\be
\label{Lehn}
\hat{c}_{1}=-\sum\limits_{n,m=1}^{\infty}\, n m p_{n+m} \dfrac{\partial^2}{\partial p_{m}\partial p_{n}}
\ee
Note, that this operator increases the cohomological degree by $2$ as it should.

\subsection{}
Now, let us consider the equivariant situation.
Let $T=({\mathbb{C}}^{\ast})^2$ be a two-dimensional torus acting on the plane by scaling coordinates:
\be
\label{tact}
(z_1,z_2) : (x,y)\mapsto (z_1 x, z_2 y)
\ee
This induces the obvious action on polynomial ideals and, therefore, on the Hilbert schemes. The equivariant cohomologies of Hilbert schemes defined by this action can be considered as two-parametric "deformation" of usual cohomologies:
$$
H_{T}^{\bullet}\Big(\coprod\limits_{n=1}^{\infty} \textrm{Hilb}_{n} \Big) \simeq  {\mathbb{C}}[p_1,p_2,...] \otimes {\mathbb{C}}[\epsilon_1, \epsilon_2]
$$
where the "deformation" parameters $\epsilon_1$ and $\epsilon_2$ have cohomological degree two: $\deg (\epsilon_1)=\deg (\epsilon_2) =2$ and generate the ring of characters of $T$. The usual cohomologies arise as a limit:
$$
H_{T}^{\bullet}\Big(\coprod\limits_{n=1}^{\infty} \textrm{Hilb}_{n} \Big) \ \ \stackrel{\epsilon_1=0, \ \ \epsilon_2=0}{\longrightarrow} \ \ H^{\bullet}\Big(\coprod\limits_{n=1}^{\infty} \textrm{Hilb}_{n} \Big)
$$
In the equivariant case, the operators of cup product \textit{are not nilpotent} because in this case for all $n>0$ the corresponding cohomologies do not vanish $H^{n}_{T}(\textrm{Hilb}_{m} )\neq 0$. \textbf{The operators of cup product becomes diagonal in the basis of equivariant classes of fixed points}  (i.e. points on $\textrm{Hilb}_{n}$ invariant under action of $T$). The first Chern class in the equivariant case takes the form~\cite{OP,InstR}:
\be
\label{fcc}
\hat{c}_{1} = \sum\limits_{m,n=1}^{\infty}\, \Big( \epsilon_1 \epsilon_2 (n+m) p_{n} p_{m} \dfrac{\partial}{\partial p_{n+m}} - n m p_{n+m} \dfrac{\partial^2}{ \partial p_n \partial p_m} \Big)+ \dfrac{\epsilon_1+\epsilon_2}{2} \sum\limits_{n-1}^{\infty} \,(n-1) n p_n \dfrac{\partial}{\partial p_n}
\ee
Note that this operator increases the degree by two, and in the limit $\epsilon_1=\epsilon_2=0$ we  obtain the classical formula of Lehn~(\ref{Lehn}).

\subsection{}
Assume that the ideal ${\cal{J}} \in \textrm{Hilb}_{n}$ is fixed under the action (\ref{tact}), then it must be generated by monomials $x^a y^{b}$ for some natural numbers $a$ and $b$. Such a monomials form a $\mathbb{C}$-basis in ${\mathbb{C}}[x,y]$ and can be represented by a table as in fig.\ref{P3}. If $x^a y^b$ is a generator of ideal, then, obviously all elements of the table above and to the right of $(a,b)$ are also in the ideal. Thus, the ideals generated by monomials with finite codimension $\dim {\mathbb{C}}[x,y]/{\cal{J}}=n$ are represented by Young diagrams (partitions) $\lambda$ with $|\lambda|=n$.
\begin{figure}[h!]
\begin{center}
\includegraphics[scale=0.6]{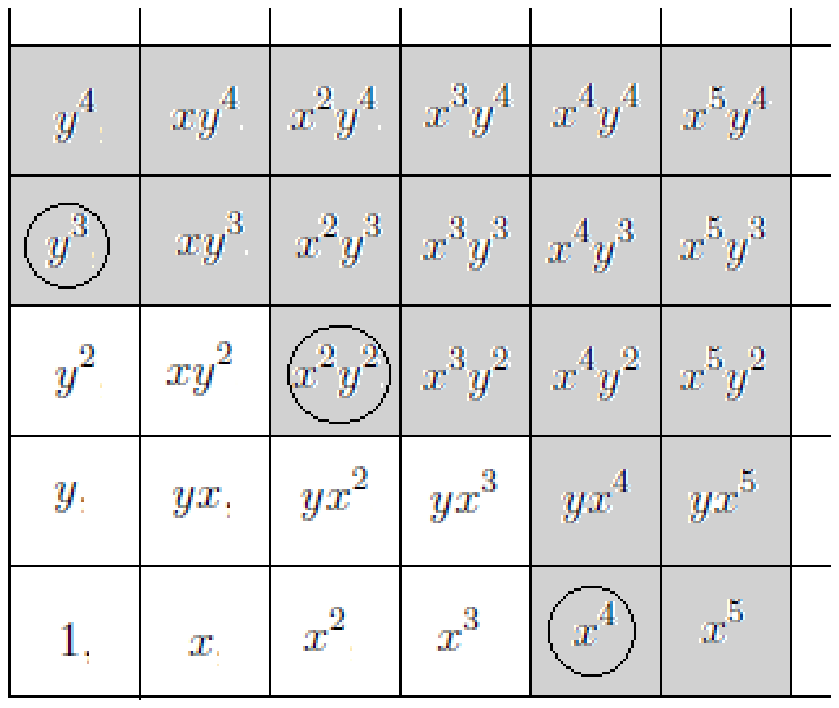}
\caption{\small{The ideal ${\cal{J}}=\langle y^3, x^2 y^2 , x^4 \rangle$ corresponds to $\lambda=[3,3,2,2]$} }
\label{P3}
\end{center}\end{figure}
For example, consider the ideal generated by monomials $y^3, x^2 y^2 , x^4$ fig.\ref{P3}. The elements of the ideal are represented by grey color in the figure above. The space ${\mathbb{C}}[x,y]/{\cal{J}}$ is 10-dimensional and spanned by monomials inside Young diagram $\lambda=[3,3,2,2]$ and are represented by white color in fig.\ref{P3}

Thus, the $T$-fixed points on $\textrm{Hilb}_{n}$ are enumerated by partitions $\lambda$ with $|\lambda|=n$.
\subsection{}
 It was noted that operator (\ref{fcc}) coincides with the hamiltonian of quantum trigonometric Calogero-Moser-Sutherland (CMS) system for infinite number of particles. This hamiltonian is known to be diagonal in the basis of Jack polynomials. \textbf{Therefore, we have a natural identification of the fixed point classes in the equivariant cohomologies with the Jack polynomials.} For example, the first several Jack polynomials $j_{\lambda}$ labeled by a partition $\lambda$ have the following form:
\be
\label{jp}
\begin{array}{c}
j_{[1]}=p_{{1}}\epsilon_{{2}}\epsilon_{{1}}; \ \ \ j_{[2]}=\epsilon_{{1}}{\epsilon_{{2}}}^{2}p_{{2}}+{\epsilon_{{1}}}^{2}{
\epsilon_{{2}}}^{2}{p_{{1}}}^{2}, \ \ \ j_{[1,1]}=\epsilon_
{{2}}{\epsilon_{{1}}}^{2} p_{{2}}+{\epsilon_{{1}}}^{2}{\epsilon_{{2}}}^{2}{p_{{1}}}^{2}\\
\\
j_{[3]}={\epsilon_{{1}}}^{3}{\epsilon_{{2}}}^{3}{p_{{1}}}^{3}+2\,\epsilon_{{1}
}{\epsilon_{{2}}}^{3}p_{{3}}+3\,{\epsilon_{{1}}}^{2}{\epsilon_{{2}}}^{
3}p_{{1}}p_{{2}},  \ \ j_{[1,1,1]}={\epsilon_{{1}}}^{3}{\epsilon_{{2}}}^{3}{p_{{1}}}^{3}+3\,p_{{1}}{
\epsilon_{{2}}}^{2}{\epsilon_{{1}}}^{3}p_{{2}}+2\,p_{{3}}\epsilon_{{2}
}{\epsilon_{{1}}}^{3} \\
\\
j_{[2,1]}={\epsilon_{{1}}}^{2}{\epsilon_{{2}}}^{2}p_{{3}}+{\epsilon_{{1}}}^{2}{
\epsilon_{{2}}}^{3}p_{{1}}p_{{2}}+p_{{1}}{\epsilon_{{2}}}^{2}{\epsilon
_{{1}}}^{3}p_{{2}}+{\epsilon_{{1}}}^{3}{\epsilon_{{2}}}^{3}{p_{{1}}}^{
3}
\end{array}
\ee
Note, that the classes of fixed points have correct degree $\deg j_{\lambda} = 4 |\lambda|=\dim_{{\mathbb{R}}}( \textrm{Hilb}_{|\lambda|})$  and respects the symmetry in the choice of generators of the torus $T \simeq{\mathbb{C}}^{\ast}_1 \times {\mathbb{C}}^{\ast}_2 \simeq {\mathbb{C}}^{\ast}_2 \times {\mathbb{C}}^{\ast}_1$:
\be \label{relt} j_{\lambda}(\epsilon_1,\epsilon_2)=j_{\lambda^{\prime}}(\epsilon_2,\epsilon_1) \ee

Let ${\cal{V}}_{\lambda}$ be the fiber of some equivariant bundle over the point $\lambda$. This space is natural $T$-module, therefore is decomposed to one-dimensional irreducibles ${\cal{V}}_{\lambda}=\oplus_{i=1}^{\dim \cal{V}} {\mathbb{C}}_{\chi_i^{\lambda}}$ with characters $\chi_i^{\lambda}$.
In the equivariant case, the eigenvalues of characteristic classes in the basis of fixed points are given by symmetric functions in $\chi_i$. For example, the eigenvalues of Chern classes are the elementary symmetric  functions in  $\chi_i^{\lambda}$:
\be
c_{k}({\cal{V}}) j_{\lambda}= e_{k}(\chi_1^{\lambda},...,\chi_{\dim \cal{V}_\lambda}^{\lambda}) j_{\lambda}
\ee
such that, for example, for the first Chern class we have:
\be
c_{1}({\cal{V}}) j_{\lambda}=(\chi_1^{\lambda}+...+\chi_{\dim \cal{V}_\lambda}^{\lambda}) j_{\lambda}
\ee
The fiber of tautological bundle over the fixed point $\lambda$ is spanned by $x^a y^b$ for $(a,b)$ inside the Young diagram $\lambda$ as in fig.\ref{P3}.
The character of  one-dimensional subspace spanned by $x^a y^b$ is obviously $a \epsilon_1+b \epsilon_2$. Thus, for the eigenvalues of CMS hamiltonian, representing the first Chern class we have:
\be
\hat{c}_{1} j_{\lambda} = \varphi_{\lambda} j_{\lambda}, \ \ \ \varphi_\lambda=\sum\limits_{(i,j) \in \lambda} (i-1)\, \epsilon_1 +(j-1) \, \epsilon_2
\ee

\subsection{}
Theoretically, the normalization of the Jack polynomials considered above is the most attractive: most of the properties and symmetries of these polynomials are obvious in this parametrization.  However, in order to make the relations to our previous papers \cite{MMS,MMSS} clear, in the present text we use conventional normalization of Jack polynomials generally accepted in the theory of matrix models. The relation between the above formulae and those used in this paper is given by change of the time variables $p_{k}\rightarrow - p_{k}/\epsilon_2$ and introduction of a new variable $\beta=-\epsilon_1/\epsilon_2$. Such that the above formulae for Jack polynomials turns to:
\be
\begin{array}{c}
j_{[1]}=p_1\ \ \
j_{[2]}={p_{{1}}}^{2}+{\dfrac {p_{{2}}}{\beta}} \ \ \
j_{[1,1]}={p_{{1}}}^{2}-p_{{2}}\\
\\
j_{[3]}={p_{{1}}}^{3}+3\,{\dfrac {p_{{1}}p_{{2}}}{\beta}}+2\,{\dfrac {p_{{3}}}{{
\beta}^{2}}} \ \
j_{[2,1]}={p_{{1}}}^{3}-{\dfrac { \left( -1+\beta \right) p_{{2}}p_{{1}}}{\beta}}
-{\dfrac {p_{{3}}}{\beta}}\ \ \
j_{[1,1,1]}={p_{{1}}}^{3}-3\,p_{{1}}p_{{2}}+2\,p_{{3}}
\end{array}
\ee
Note, also, that we normalize  $j_{\lambda}$ such that the coefficient of $p_{1}^{|\lambda|}$ is trivial.

\noindent
The first Chern class in this normalization takes the form:
\be
\hat{c}_{1}=\dfrac{1}{2} \sum\limits_{n,m=1}^{\infty}\, \beta (n+m) p_m p_n \dfrac{\partial}{ \partial p_{n+m}} + n m p_{n+m} \dfrac{\partial^2}{ \partial p_n \partial p_m} -\dfrac{1-\beta}{2} \sum\limits_{n=1}^{\infty} \, (n-1) n p_n \dfrac{\partial}{\partial p_n }
\ee
such that we have:
\be
\hat{c}_{1}j_{\lambda} = \varphi_{\lambda}(0) j_{\lambda}, \ \ \ \varphi_{\lambda}(0)= \sum\limits_{(i,j)\in \lambda} \, \Big( (i-1) -(j-1)\beta \Big)
\ee
with $\phi_{\lambda}(u)$ defined in (\ref{phifun}). Relation (\ref{relt}) turns to:
\be
\label{recipr}
j^{\beta}_{\lambda}(-p_{k}/\beta)=j^{1/\beta}_{\lambda^{\prime}}(p_k)\, \beta^{-2  |\lambda|}
\ee

\subsection{}
The Fock space has a natural scalar product,  induced from topology of Hilbert schemes, such that the classes of the fixed points are orthogonal. It is defined as follows: let us consider the inclusion of a fixed point $\lambda$ to $\textrm{Hilb}_{n}$, and let $i^{\ast}_{\lambda}$ be the corresponding pullback (restriction) map:
\be
 \textrm{pt} \stackrel{i_{\lambda} }{\longrightarrow} \textrm{Hilb}_{n}, \ \ \ \ H^{\bullet}_{T}( \textrm{Hilb}_{n} )  \stackrel{i_{\lambda}^{\ast} }{\longrightarrow} H^{\bullet}_{T}(\textrm{pt})\simeq {\mathbb{C}}[\epsilon_1, \epsilon_2]
\ee
Then, we can define the scalar product in the basis of fixed points $j_{\lambda}$ as:
\be
\label{spd}
\langle j_{\lambda}, j_{\mu} \rangle \stackrel{def}{=}i^{\ast}_{\lambda}( j_{\mu} )=\delta_{\lambda \mu} \, e(T_{\lambda} \textrm{Hilb}_{n})
\ee
The tangent space  $T_{\lambda} \textrm{Hilb}_{n}$ is a natural representation of the torus $T$, thus is decomposed to linear irreducibles with some characters $\chi_k$: $ T_{\lambda} \textrm{Hilb}_{n}= \oplus_{k=1}^{2n} {\mathbb{C}}_{\chi_k}$. The Euler class or determinant of this representation is defined as the product of all characters encountered in it:
$$
e(T_{\lambda} \textrm{Hilb}_{n}) = \prod\limits_{k=1}^{2 n} \chi_k
$$
It is well known, that the tangent space at the ideal ${\cal{J}}_{\lambda}$ is given by the ext functor $T_{\lambda} \textrm{Hilb}_{n}\simeq \textrm{Ext}^{1}({\cal{J}}_{\lambda},{\cal{J}}_{\lambda})$ thus, the corresponding character decomposition can be computed explicitly from some free resolution of ${\cal{J}}_{\lambda}$ as for example it was done in \cite{Carl}:
$$
e(T_{\lambda} \textrm{Hilb}_{n})=\prod\limits_{\Box \in \lambda}\,\Big( (a_{\lambda}(\Box)+1) \epsilon_2 -l_{\lambda}(\Box) \epsilon_1 \Big) \Big( (l_{\lambda}(\Box)+1) \epsilon_1 -a_{\lambda}(\Box) \epsilon_2 \Big) = e_{\lambda,\lambda}(0)
$$
where the functions $e_{\lambda,\mu}$ and $a_{\lambda}$, $l_{\lambda}$ are defined in (\ref{efun}) and (\ref{armleg}). Thus, the Jack polynomials normalized as above form an orthogonal basis in the Fock space and have norms given by $e_{\lambda,\lambda}(0)$:
\be
\langle j_{\lambda}(p_k), j_{\mu}(p_k) \rangle= e_{\lambda,\lambda}(0)
\ee
(here we use that $e_{\lambda,\mu}(0)=0$ for $\lambda\neq \mu$).  Note, that this scalar product is equivalent to the standard one defined as:
\be
\label{scpr}
\langle f_{\lambda}(p_n), g_{\mu}(p_n) \rangle = \left.f_{\lambda}\Big( \frac{n}{\beta} \, \dfrac{\partial}{\partial p_n} \Big) g_{\mu}(p_n)\right|_{p_k=0}
\ee

\subsection{\label{unit}}
An important element in the present text is an identity operator ${\mathbb{{I}}}(p_k,q_k)$ defined by this scalar product as follows:
\be
\Big\langle {\mathbb{{I}}}(p_k,q_k) , f(q_k)   \Big\rangle = f(p_k) , \ \ \ \forall f(q_k)
\ee
From (\ref{scpr}) it is obvious that explicitly this elements is given by:
\be
 {\mathbb{{I}}}(p_k,q_k)=\exp\Big(\beta\sum\limits_{n=1}^{\infty} \, \frac{p_n q_n}{n} \Big)
\ee
indeed from (\ref{scpr}) we have:
$$
\Big\langle {\mathbb{{I}}}(p_k,q_k) , f(q_k)   \Big\rangle= \left.\exp\Big(\sum\limits_{n=1}^{\infty} \, p_n \dfrac{\partial}{\partial q_n} \Big) f(q_k)\right|_{q_k=0}=\left.f(q_k+p_k)\right|_{q_k=0} = f(p_k)
$$
At the same time the unit in equivariant cohomologies  have the following (obvious from (\ref{spd})) expansion in the basis of fixed points
\be
{\mathbb{{I}}}(p_k,q_k)=\sum\limits_{\lambda}\, \dfrac{j_{\lambda}(p_k) j_{\lambda}(q_k)}{e(T_{\lambda} \textrm{Hilb}_{n})}=\sum\limits_{\lambda}\, \dfrac{j_{\lambda}(p_k) j_{\lambda}(q_k)}{e_{\lambda,\lambda}(0)}
\ee
This way, we obtain simple geometrical interpretation of the Cauchy completeness identity:
\be
\label{Jcauchy}
\sum\limits_{\lambda}\, \dfrac{j_{\lambda}(p_k) j_{\lambda}(q_k)}{e_{\lambda,\lambda}(0)}=\exp\Big(\beta\sum\limits_{n=1}^{\infty} \, \frac{p_n q_n}{n} \Big)
\ee

\section{Appendix B: generalized Jack polynomials and instanton moduli spaces \label{apb}}

In this section, similarly to the previous one, we consider the geometry of moduli space of $GL(N)$ instantons on $4$-sphere. We will work, actually with some regularization of this space provided by framed sheaves on ${\mathbb{P}}^2$.
\subsection{}

Let denote by ${\cal{M}}_{n,r}$ the moduli space of framed sheaves ${\cal{S}}$ on ${\mathbb{P}}^2$ with fixed Chern class $c_2({\cal{S}})=n$ (this is topological charge of instantons) and rank $r$  ( i.e. $SU(r)$ instantons ).
This space is a natural generalization of the Hilbert schemes to arbitrary rank:
\be
{\cal{M}}_{n,1} \simeq  \textrm{Hilb}_{n}
\ee
The group $\textrm{GL}(2)\times \textrm{GL}(r)$ acts naturally on ${\cal{M}}_{n,r}$. The first factor $\textrm{GL}(2)$ rotates the plane ${\mathbb{C}}^2 \subset {\mathbb{P}}^2$ keeping fixed the infinity line, and $GL(r)$ acts as the gauge group. Let denote by $A\simeq({\mathbb{C}}^{\ast})^{r}$ the torus of the gauge group, such that the total torus acting on the moduli space is $T\simeq ({\mathbb{C}}^{\ast})^{2} \times A $ where the first factor is the torus of $GL(2)$. Thus, the ring of characters of torus $T$ has the form:
\be
H_{T}^{\bullet}(\textrm{pt}) = {\mathbb{C}}[\epsilon_1,\epsilon_2,u_1,...,u_r]
\ee
where $\epsilon_1$ and $\epsilon_2$ correspond to the two-dimensional factor in $T$ and $u_i$ are the characters of $A$.
The set of the fixed points in the moduli space under the action of subtorus $A$ has the following form:
\be
\label{Hsd}
{\cal{M}}_{n,r}^{A}= \coprod\limits_{n_1+...+n_r=n} \, \textrm{Hilb}_{n_1} \times ... \times \textrm{Hilb}_{n_r}
\ee
this means simply, that the only  gauge fields invariant under the adjoint action of torus $A\subset T$ are represented by the diagonal matrices, i.e. split to $r$ $U(1)$ factors. Thus, as follows from the \textbf{appendix A}, the fixed points of the total torus are given by $r$-tuples of partitions:
\be
\label{tfp}
{\cal{M}}_{n,r}^{T}=\{ (\lambda_1,...,\lambda_r) : |\lambda_1|+...+|\lambda_r|=n \}
\ee
i.e. these are the  points in (\ref{Hsd}) fixed under the action of first two dimensional factor in $T$. From the standard isomorphism in equivariant cohomologies that identifies the cohomologies of total space with cohomologies of fixed points we obtain:
\be
\label{isomc}
H^{\bullet}_{T}({\cal{M}}_{n,r})\simeq H^{\bullet}_{T}({\cal{M}}^A_{n,r})
\ee
As in the previous section we consider all moduli spaces for different $n$ at one go: let us consider the "composite" space ${\cal{M}}(r)=\coprod_n {\cal{M}}_{n,r}$. From, (\ref{Hsd}) and (\ref{isomc}) is follows that:
\be
\label{fsp}
H^{\bullet}_{T}({\cal{M}}(r))=\underbrace{{\cal{F}} \otimes ...\otimes {\cal{F}}}_{r-\textrm{times}}  \otimes {\mathbb{C}}(\epsilon_1,\epsilon_2,u_1,...,u_r)
\ee
Thus, the cohomologies of rank $=r$ instantons are given by tensor product of $r$ boson Fock spaces and the cohomology classes can be represented  by  polynomials in $r$ -time variables $p^{(m)}_{k}$ $m=1...r$ with coefficients which are rational functions in in $r+2$ equivariant parameters $\epsilon_1,\epsilon_2,u_1,...,u_r$.

\subsection{}

The most interesting to us are the polynomials corresponding to $T$-fixed points. As it was already noted above (\ref{tfp}), these polynomials labeled by $r$-tuples of partitions, and are some natural generalizations of the Jack polynomials to arbitrary rank. We denote them $J_{\lambda_1,...,\lambda_r}$ and call them \textrm{generalized Jack polynomials}.
 As in the case of Hilbert schemes, we can define the classes of the fixed points as eigenvectors of Chern classes of tautological bundle ${\cal{V}}$ over the moduli space. The fibers of ${\cal{V}}$ over a sheaf ${\cal{S}} \in {\cal{M}}_{n,r}$ is defines as a space of global sections:
\be
\label{fiber}
\left. {\cal{V}}\right|_{{\cal{S}}}=H^{1}({\mathbb{P}}^2,{{\cal{S}}})
\ee
Under the identification (\ref{fsp}), the action of Chern classes are represented by a set of commuting operators (Hamiltonians). The resulting integrable system  generalizes the trigonometric CMS system to arbitrary number of time variables. The explicit formula for the first Chern class is calculated, for example, in \cite{Ok} and have to following form:
\be
\label{hamil}
c_1= \sum\limits_{k=1}^{r}\, H_{k} + \sum\limits_{k_1<k_2} \, H_{k_1,k_2}
\ee
where $H_{k}$ in the first sum is the shifted CMS Hamiltonian acting in the $k$-th component of the tensor product (\ref{fsp}):
\be
\label{dir}
H_{k}=\dfrac{1}{2} \sum\limits_{n,m=1}^{\infty}\, \beta (n+m) p_m^{(k)} p_n^{(k)} \dfrac{\partial}{ \partial p_{n+m}^{(k)}} + n m p_{n+m}^{(k)} \dfrac{\partial^2}{ \partial p_n^{(k)} \partial p_m^{(k)}} +\dfrac{1}{2} \sum\limits_{n=1}^{\infty} \, \Big(2 u_k+(\beta-1)(n-1)\Big) n p_n^{(k)} \dfrac{\partial}{\partial p_n^{(k)} }
\ee
and $ H_{k_1,k_2}$ is the "mixing term" acting in $k_1$-th and $k_2$-th tensor component:
\be
\label{mix}
H_{k_1,k_2}=(1-\beta)\sum\limits_{n=1}^{\infty} n^2 p_{n}^{(k_1)} \dfrac{\partial}{\partial p_n^{(k_2)} }
\ee
Note, that at $\beta=1$ this mixing term disappears, and we obtain a sum of noninteracting CMS hamiltonians. Thus, the eigenvectors at $\beta=1$ are represented by a product of $r$ Schur functions $J_{\lambda_1,...,\lambda_r}=s_{\lambda_1}...s_{\lambda_r}$. In general, the mixing term is not zero and there is no such a factorization for eigenfunctions. The classes of the fixed points can be defined (up to a multiple) as eigenfunctions of hamiltonian (\ref{hamil}), that have a proper limit for $\beta=1$ (i.e. factorize to corresponding product of Schur functions).
\subsection{}
The eigenvalues of the first Chern class are given by the character of fiber (\ref{fiber}) at the fixed point $(\lambda_1,...,\lambda_r)$. The sheaf representing the fixed point is a direct sum of rank one ideal sheaves ${\cal{S}}_{\lambda_1,...,\lambda_r}\simeq{\cal{J}}^{u_1}_{\lambda_1}\oplus...\oplus {\cal{J}}_{\lambda_r}^{u_r}$. Thus the cohomology is a direct sum of rank one terms:
$$
H^{1}({\mathbb{P}}^2,{\cal{S}}_{\lambda_1,...,\lambda_r}) =  \bigoplus\limits_{i=1}^{r} H^{1}({\mathbb{P}}^2,{\cal{J}}_{\lambda_i}^{u_i})
$$
and the character, of this space is a sum of individual rank one characters, thus for eigenvalues we obtain:
\be
\hat{c_{1}} J_{\lambda_1,...,\lambda_r}= \Big(  \varphi_{\lambda_1}(u_1)+...+\varphi_{\lambda_r}(u_r)   \Big)\, J_{\lambda_1,...,\lambda_r}
\ee
with
\be
\label{phifun}
\varphi_{\lambda}(u)= \sum\limits_{(i,j)\in \lambda} \, \Big(u + (i-1) +(j-1)\beta \Big)
\ee
This is also obvious from the form of hamiltonian (\ref{hamil}): it is a sum of CMS hamiltonians, plus locally nilpotent mixing term. The additional  nilpotent term, obviously can not change the eigenvalues, and thus they are the sum of eigenvalues of $r$ - CMS operators.  The extra $u$ -dependence comes from the term:
$$
D_u=u \sum\limits_{n=1}^{\infty} n p_n \dfrac{\partial}{\partial p_n }
$$
which simply counts the degree of the polynomial:
\be
D_u j_{\lambda}=|\lambda| u  \, j_{\lambda}
\ee
\subsection{}

The important difference of this hamiltonian from one Hilbert schemes (rank one) is that it is not self-adjoint, with respect to a scalar product (\ref{scpr}). Indeed, under the conjugation:
\be
(p_{n}^{(k)})^{\ast}=-\dfrac{n}{\beta} \dfrac{\partial}{\partial p_{n}^{(k)}}
\ee
the operators (\ref{dir}) is self-adjoint, but the mixing term (\ref{mix}) is not. We denote by $J^{\ast}_{\lambda_1,...,\lambda_r}$ the eigenvectors of the adjoint operator $H^{\ast}$. Note, that under the conjugation, the mixing term transforms such that $u_{i}\rightarrow u_{r-i+1}$ and $p_k^{(i)}\rightarrow p_{k}^{r-i+1}$, thus the relation among generalized Jack polynomials and its dual is very simple:
\be
\label{drd}
J_{\lambda_1,...,\lambda_r}^{u_1,...,u_r}(p^{(1)},...,p^{(r)})={J^{\ast}}_{\lambda_r,...,\lambda_1}^{u_r,...,u_1}(p^{(r)},...,p^{(1)})
\ee
We defined the functions $J_{\lambda_1,...,\lambda_r}$ up to a multiple. This multiple is fixed by (\ref{drd}) together with a normalization:
\be
\label{normcond}
\langle J_{\lambda_1,...,\lambda_r}, J^{\ast}_{\mu_1,...,\mu_r} \rangle =\delta_{\lambda_1,\mu_1} ... \delta_{\lambda_r,\mu_r} e( T_{\lambda_1,...\lambda_r} {\cal{M}}_{r,n})
\ee
which generalizes (\ref{spd}) to arbitrary rank.   To compute the character of the tangent space $T_{\lambda_1...\lambda_r} {\cal{M}}_{r,n}$, we should note that the sheaf  corresponding to the fixed point is the sum of ideal sheaves: ${\cal{S}}_{\lambda_1,...,\lambda_r}\simeq{\cal{J}}_{\lambda_1}\oplus...\oplus {\cal{J}}_{\lambda_r}$, thus:
$$
T_{\cal{S}} {\cal{M}}_{r,n} \simeq \textrm{Ext}^{1}({\cal{S}},{\cal{S}}) \simeq \bigoplus\limits_{1 \leq i,j \leq r} \textrm{Ext}^{1}({\cal{J}}_{\lambda_i},{\cal{J}}_{\lambda_j})
$$
what gives the following explicit expression for the character \cite{Carl}:
$$
e( T_{\lambda_1,...\lambda_r} {\cal{M}}_{r,n})  =\prod\limits_{i,j=1}^{r} \, e_{\lambda_i,\mu_{j}}(u_i-u_j)
$$
with function $e_{\lambda_i,\lambda_j}$ given explicitly by:
\be\label{efun}
e_{\lambda,\mu}(u)=\dfrac{(-1)^{|\lambda|}}{\beta^{|\lambda|+|\mu|}}\, \prod\limits_{\Box\in \lambda} ( u+a_{\lambda}(\Box)+1+\beta l_{\mu}(\Box) ) \prod\limits_{\Box \in \mu} \,( u-a_{\mu}(\Box)-\beta l_{\lambda}(\Box)-\beta )
\ee
where $a_{\lambda}(\Box)$ and $l_{\lambda}(\Box)$ are the standard arm and leg length of the box $\Box$ in the Young diagram $\lambda$. If the box has coordinates $(i,j)$, then the corresponding functions are defined as:
\be
\label{armleg}
a_{\lambda}(\Box)=\lambda_{i}-j, \ \ \ l_{\lambda}(\Box)=\lambda^{\prime}_{j}-i
\ee
From the definition its is clear that $e_{\lambda,\mu}(0)=0$ of  $\mu \neq \lambda$. Thus, the condition (\ref{normcond}) takes the following form:
\be
\langle J_{\lambda_1,...,\lambda_r}, J^{\ast}_{\mu_1,...,\mu_r} \rangle =\prod\limits_{i,j=1}^{r} \, e_{\lambda_i,\mu_{j}}(u_i-u_j)
\ee
\subsection{}
Repeating the "unit"argument of section \ref{unit}, we obtain the following Cauchy identity for the generalized Jack functions:
\be
\label{GJcauchy}
{\mathbb{{I}}}=\sum\limits_{\lambda_1,...,\lambda_r} \, \dfrac{ J_{\lambda_1,...,\lambda_r}(p^{(1)}_{k_{1}},...,p^{(r)}_{k_{r}})\, J^{\ast}_{\lambda_1,...,\lambda_r}(q^{(1)}_{k_{1}},...,q^{(r)}_{k_{r}}) }{\prod\limits_{i,j=1}^{r} \, e_{\lambda_i,\mu_{j}}(u_i-u_j)} = \exp\Big(  \beta \sum\limits_{n=1}^{\infty} \dfrac{p^{(1)}_n q^{(1)}_n+...+p^{(r)}_n q^{(r)}_n}{n} \Big)
\ee
Note, also, that the analog of the relation (\ref{recipr}) for generalized Jack polynomials gives:
\be
J^{\beta}_{\lambda_1,...,\lambda_r}(u_i,-p^{(i)}_k/\beta)=J^{1/\beta}_{\lambda_1^{\prime},...,\lambda_r^{\prime}}(-u_i/\beta,p^{(i)}_k)\, \beta^{-2 |\lambda_1|-...-2 |\lambda_r|}
\ee
To this end, we give explicit expressions for the first several generalized Jack polynomials in the case of $r=2$, which are used in this paper. In the accepted normalization these polynomials read:

\be
\label{fpc}
\begin{array}{l}
J_{[1],[]} (a,x_i,y_i)=-2\,{\frac {{\it a}\,x_{{1}}}{\beta}}-{\frac { \left( 1-\beta
 \right) y_{{1}}}{\beta}}\\
J_{[],[1]} (a,x_i,y_i)={\frac { \left( 2\,{\it a}-1+\beta \right) y_{{1}}}{\beta}}\\
J_{[2],[]} (a,x_i,y_i)=
{\frac { \left( 4\,\beta\,{{\it a}}^{2}+2\,\beta\,{\it a} \right) {x
_{{1}}}^{2}}{{\beta}^{3}}}+{\frac { \left( 4\,\beta\,{\it a}-4\,{
\beta}^{2}{\it a} \right) y_{{1}}x_{{1}}}{{\beta}^{3}}}+{\frac {
 \left( 2\,{\it a}+4\,{{\it a}}^{2} \right) x_{{2}}}{{\beta}^{3}}}+{
\frac { \left( {\beta}^{3}+2\,\beta-3\,{\beta}^{2} \right) {y_{{1}}}^{
2}}{{\beta}^{3}}}+{\frac { \left( 2+4\,{\it a}-3\,\beta-4\,\beta\,{
\it a}+{\beta}^{2} \right) y_{{2}}}{{\beta}^{3}}}\\
J_{[1,1],[] } (a,x_i,y_i)={\frac { \left( 4\,{{\it a}}^{2}-2\,\beta\,{\it a} \right) {x_{{1}}}
^{2}}{{\beta}^{2}}}+{\frac { \left( 4\,{\it a}-4\,\beta\,{\it a}
 \right) y_{{1}}x_{{1}}}{{\beta}^{2}}}+{\frac { \left( 2\,\beta\,{\it
a}-4\,{{\it a}}^{2} \right) x_{{2}}}{{\beta}^{2}}}+{\frac { \left( -
3\,\beta+1+2\,{\beta}^{2} \right) {y_{{1}}}^{2}}{{\beta}^{2}}}+{\frac
{ \left( -1-4\,{\it a}+3\,\beta+4\,\beta\,{\it a}-2\,{\beta}^{2}
 \right) y_{{2}}}{{\beta}^{2}}}\\
J_{[],[2] }(a,x_i,y_i)={\frac { \left( 2\,{\it a}-2+\beta \right)  \left( 2\,{\it a}-1+
\beta \right) {y_{{1}}}^{2}}{{\beta}^{2}}}+{\frac {y_{{2}} \left( 2\,{
\it a}-2+\beta \right)  \left( 2\,{\it a}-1+\beta \right) }{{\beta}^
{3}}}\\
J_{[],[1,1]} (a,x_i,y_i)={\frac { \left( 2\,{\it a}-1+2\,\beta \right)  \left( 2\,{\it a}-1+
\beta \right) {y_{{1}}}^{2}}{{\beta}^{2}}}-{\frac {y_{{2}} \left( 2\,{
\it a}-1+2\,\beta \right)  \left( 2\,{\it a}-1+\beta \right) }{{
\beta}^{2}}}\\
J_{[1],[1]} (a,x_i,y_i)=-{\frac { \left( 2\,\beta\,{\it a}-2\,{\it a}+4\,{{\it a}}^{2}-
\beta \right) y_{{1}}x_{{1}}}{{\beta}^{2}}}-{\frac { \left( 2\,{\it a
}-1-2\,\beta\,{\it a}+2\,\beta-{\beta}^{2} \right) {y_{{1}}}^{2}}{{
\beta}^{2}}}-{\frac { \left( 1-\beta \right) y_{{2}}}{{\beta}^{2}}}
 \end{array}
\ee
the dual basis have the form:

\be
\begin{array}{l}
J^{\ast}_{[1],[]} (a,x_i,y_i)=-{\frac { \left( -\beta+2\,{\it a}+1 \right) x_{{1}}}{\beta}}\\
J^{\ast}_{[],[1]} (a,x_i,y_i)={\frac { \left( -1+\beta \right) x_{{1}}}{\beta}}+2\,{\frac {{\it a}
\,y_{{1}}}{\beta}}\\
J^{\ast}_{[2],[]} (a,x_i,y_i)={\frac { \left( -\beta+2\,{\it a}+1 \right)  \left( 2\,{\it a}+2-
\beta \right) {x_{{1}}}^{2}}{{\beta}^{2}}}+{\frac { \left( -\beta+2\,{
\it a}+1 \right)  \left( 2\,{\it a}+2-\beta \right) x_{{2}}}{{\beta}
^{3}}}\\
J^{\ast}_{[1,1],[]} (a,x_i,y_i)={\frac { \left( 2\,{\it a}+1-2\,\beta \right)  \left( -\beta+2\,{\it
a}+1 \right) {x_{{1}}}^{2}}{{\beta}^{2}}}-{\frac { \left( 2\,{\it a}
+1-2\,\beta \right)  \left( -\beta+2\,{\it a}+1 \right) x_{{2}}}{{
\beta}^{2}}}\\
J^{\ast}_{[],[2]} (a,x_i,y_i)={\frac { \left( {\beta}^{3}+2\,\beta-3\,{\beta}^{2} \right) {x_{{1}}}^
{2}}{{\beta}^{3}}}+{\frac { \left( 4\,{\beta}^{2}{\it a}-4\,\beta\,{
\it a} \right) y_{{1}}x_{{1}}}{{\beta}^{3}}}+{\frac { \left( 2+4\,
\beta\,{\it a}-3\,\beta-4\,{\it a}+{\beta}^{2} \right) x_{{2}}}{{
\beta}^{3}}}+{\frac { \left( 4\,\beta\,{{\it a}}^{2}-2\,\beta\,{\it
a} \right) {y_{{1}}}^{2}}{{\beta}^{3}}}+{\frac { \left( -2\,{\it a}+
4\,{{\it a}}^{2} \right) y_{{2}}}{{\beta}^{3}}}\\
J^{\ast}_{[],[1,1]} (a,x_i,y_i)={\frac { \left( -3\,\beta+1+2\,{\beta}^{2} \right) {x_{{1}}}^{2}}{{
\beta}^{2}}}+{\frac { \left( -4\,{\it a}+4\,\beta\,{\it a} \right) y
_{{1}}x_{{1}}}{{\beta}^{2}}}+{\frac { \left( -1+4\,{\it a}-4\,\beta\,
{\it a}+3\,\beta-2\,{\beta}^{2} \right) x_{{2}}}{{\beta}^{2}}}+{
\frac { \left( 2\,\beta\,{\it a}+4\,{{\it a}}^{2} \right) {y_{{1}}}^
{2}}{{\beta}^{2}}}+{\frac { \left( -4\,{{\it a}}^{2}-2\,\beta\,{\it
a} \right) y_{{2}}}{{\beta}^{2}}}\\
J^{\ast}_{[1],[1]} (a,x_i,y_i)=-{\frac { \left( -1-2\,{\it a}-{\beta}^{2}+2\,\beta\,{\it a}+2\,
\beta \right) {x_{{1}}}^{2}}{{\beta}^{2}}}-{\frac { \left( -2\,\beta\,
{\it a}+2\,{\it a}-\beta+4\,{{\it a}}^{2} \right) y_{{1}}x_{{1}}}{{
\beta}^{2}}}-{\frac { \left( 1-\beta \right) x_{{2}}}{{\beta}^{2}}}
\end{array}
\ee
where $p_{k}^{(1)}=x_k$ and $p_{k}^{(2)}=y_{k}$ and we identify $u_1=-u_2=a$ such that the polynomials depend only on $u_1-u_2=2 a$.

\section*{Acknowledgements}
We are grateful to A.Okounkov, M. McBreen, A. Negut, V.Alba, K.Gimre and D.~Galakhov  for helpful discussions and Y. Matsuo  for important comments on the text. Our work was partly supported by Ministry of Education and Science of the Russian Federation under contract 8498, the Brazil National Counsel of Scientific and Technological Development, by NSh-3349.2012.2,
RFBR grants  12-02-00594, 13-02-00478 , 12-01-00482, by joint grants 12-02-92108-Yaf, 13-02-91371-ST, 14-01-93004-Viet, by leading young scientific groups RFBR 12-01-33071 mol-a-ved.

\end{small}
\end{document}